\documentclass[journal=jacsat,manuscript=article]{achemso}

\usepackage{chemformula} 
\usepackage[T1]{fontenc} 
\usepackage{siunitx}
\usepackage{hyperref}

\newcommand*{\citen}[1]{%
  \begingroup
    \romannumeral-`\x 
    \setcitestyle{numbers}%
    \cite{#1}%
  \endgroup   
}

\author{Zhenyang Xiao}
\affiliation[University of Notre Dame]
{Department of Electrical Engineering, University of Notre Dame, Notre Dame}
\email{zxiao3@nd.edu}
\author{Jiashu Wang}
\affiliation[University of Notre Dame]
{Department of Physics, University of Notre Dame, Notre Dame}
\author{Xinyu Liu}
\affiliation[University of Notre Dame]
{Department of Physics, University of Notre Dame, Notre Dame}
\author{Badih Assaf}
\affiliation[University of Notre Dame]
{Department of Physics, University of Notre Dame, Notre Dame}
\author{David Burghoff}
\affiliation[University of Notre Dame]
{Department of Electrical Engineering, University of Notre Dame, Notre Dame}

\title[An \textsf{achemso} demo]
  {Optical-pump terahertz-probe spectroscopy of the topological crystalline insulator Pb$_{1-x}$Sn$_{x}$Se through the topological phase transition}

\abbreviations{IR,NMR,UV}
\keywords{American Chemical Society, \LaTeX}

\begin{document}

\begin{abstract}
Topological crystalline insulators---topological insulators whose properties are guaranteed by crystalline symmetry---can potentially provide a promising platform for terahertz optoelectronic devices, as their properties can be tuned on demand when layered in heterostructures. We perform the first optical-pump terahertz-probe spectroscopy of topological crystalline insulators, using them to study the dynamics of Pb$_{1-x}$Sn$_{x}$Se as a function of temperature. At low temperatures, excitation of Dirac fermions leads to an increase in terahertz transmission; from this negative photoconductivity, the intrasubband relaxation rate of 6 ps is extracted. At high temperatures where only massive fermions exist, the free-carrier losses induced by the pump reduce the terahertz transmission for the duration of the 27 ps interband lifetime. Both effects are present at temperatures near the topological-to-trivial transition. Our experimental observations provide critical details for potential applications of Pb$_{1-x}$Sn$_{x}$Se and provide a direct measurement of the topological character of Pb$_{1-x}$Sn$_{x}$Se heterostructures.
\end{abstract}

\section{Introduction}
Recent progress in topological insulators (TIs) have pointed to them as promising platforms for terahertz and far-infrared heterostructure devices\cite{futopological2011, zhangtopological2010, krizman2018tunable}. TIs are novel materials with insulating bulk bandgaps and gapless surface states, which are promising platforms for devices with novel functionalities\cite{hasan2010colloquium,qi2011topological,hasan2011three}. TI material can potentially be protected against defects and can have tunable inverted bandgaps. There are two kinds of TI materials, the canonical topological materials and the topological crystalline insulators (TCI). The canonical topological materials are narrow-gap semiconductors with an inverted band structure, such as bismuth or antimony chalcogenides\cite{wozny2020electron}. As for TCIs, they possess a crystalline symmetry that guarantees the existence of an even number of surface states\cite{hsieh2012topological}. As one of the well-known TCI materials, lead-tin-selenide (Pb$_{1-x}$Sn$_{x}$Se) has recently been reported as an interesting candidate for terahertz optoelectronic heterostructures, as its properties can be tuned using either the Sn concentration or the temperature \cite{krizman2018tunable}. In addition, certain alloy fractions remain inverted even at room temperature. In principle, this could lead towards the creation of engineered topological states, whose wavefunctions are designed to achieve a particular optical or electrical response. For example, one can imagine terahertz sources\cite{bosco2019thermoelectrically,kainz2019thermoelectric,dhillon20172017} or detectors\cite{palaferri2018noise,micheletti2021regenerative} that operate at higher temperatures by relying on states that are protected from scattering by their topological nature, not by a reduction in overlap. However, there is still a long way to go towards making optoelectronic devices out of these novel materials. Understanding the carrier dynamics and photoconductivity is the first obstacle for creating optoelectronic devices.

Optical-pump terahertz-probe spectroscopy (OPTP) and time-resolved angle resolved photoemission spectroscopy (tr-ARPES) are common methods being used in the TI dynamics study\cite{hsieh2011selective,kumar2011spatially,sobota2012ultrafast,wang2012measurement}. Tr-ARPES is a powerful tool for studying non-equilibrium quantum states, as it enables imaging of electron dynamics of an optically excited state directly in momentum space\cite{hajlaoui2013time, christiansen2019theory}. However, the bulk band structure of TIs measured in this method is largely masked by the topological surface states. In contrast, terahertz time-domain spectroscopy (THz-TDS) and ultrafast optical-pump terahertz-probe spectroscopy (OPTP) can directly interrogate the dynamics of free carriers by manipulating the photoinduced transmittance. Since light penetrates into the TI material, both bulk and surface carriers’ information can be obtained. They are particularly useful for investigations of Dirac materials, as has been demonstrated in studies of the photoinduced carrier dynamics in graphene and Bi$_2$Se$_3$.\cite{shicontrolling2014,hafez2017effects,jnawaliobservation2013,lunegative2017,yangLightControlSurface2020}. OPTP systems can provide a time-resolved measurement for studying the photoexcited carriers generated by ultrafast infrared pulses, allowing the carrier dynamics related to the photoexcitation and photorelaxation to both be extracted. As an important 2D material, graphene has been extensively studied by OPTP measurements, and anomalous negative photoinduced conductivity has been found. This negative photoconductivity was first observed in graphene monolayers produced by chemical vapor deposition (CVD) methods, and it was subsequently observed in Bi$_2$Se$_3$ thin films as well\cite{valdesaguilartime-resolved2015,sim2014ultrafast}.

In this work, we study the carrier dynamics of the TCI Pb$_{1-x}$Sn$_{x}$Se as a function of temperature using optical-pump terahertz-probe spectroscopy. At low temperatures (T=29 K), we observe negative photoconductivity introduced by the photoexcitation, similar to what has been observed in Bi$_2$Se$_3$ and graphene. A decay time of about 6 ps is observed for the hot Dirac electrons causing the negative conductivity. Both positive and negative photoconductivity are observed at different pump delay positions at moderate temperatures (T=96, 163, and 230 K), which can be explained by the coexistence of the Dirac fermions from the gapless surface states and the massive fermions from the bulk states. At room temperature (T=297 K), only massive fermions are present, as the gapless Dirac cone is gapped, and the induced photoconductivity decays at a rate of about 27 ps. These observations provide critical details for potential applications of Pb$_{1-x}$Sn$_{x}$Se and provide a direct measurement of the topological character of topological surface states.

\section{Results and discussion}
 For this work, we grew a 50 nm quantum well of Pb$_{1-x}$Sn$_{x}$Se of BaF$_2$ (111) with a 5 nm EuSe capping layer to protect from oxidation. Pb$_{1-x}$Sn$_{x}$Se is a well-known TCI material that can exhibit a topological transition from a trivial insulator to TCI under modification of chemical composition, temperature, or pressure \cite{wozny2020electron}, while EuSe is topologically trivial. The layers are grown following the procedure described in Ref. \citen{wang2020weak}, directly on the BaF$_2$ surface, without the buffer layer stack. The band structure parameters of bulk Pb$_{1-x}$Sn$_{x}$Se has been investigated at different Sn compositions and temperatures, with a band gap $E_{g}$ that can be approximately expressed by\cite{krizman2018dirac}
\begin{align} \label{e1}
   E_g (x,T)=&E_g(x,4.2)+2\left(\sqrt{22.5^2+\left(0.32\,\si{K^{-1}}
   \right)^2(T-4.2 \,\si{K})^2} -22.5 \right)(\si{meV}) \\
   &E_g (x,4.2) =2\left(72-710x+2440x^2-4750x^3\right)(\si{meV}), \nonumber
\end{align}
 where x is the composition fraction of the Sn, T is the temperature, and $E_g(x,4.2)$ is the energy gap at 4.2 K. Equation \ref{e1} is most accurate when $0<x<0.3$ and 4.2 K<T<200 K. The Sn content of the films can deduced from X-ray diffraction measurements using Vegard’s law as done in Ref. \citen{krizman2018dirac}. We find x=0.21$\pm$0.1. We approximate the value of the energy gap at 4.2 K to be $E_g=-30\pm5 \si{meV}$ from Equation \ref{e1}. Hall measurements indicate that the samples are p-doped at low temperature with an estimated hole concentration of $7.8 \times 10^{18}\,\si{ cm}^{-3}$.
 
 The band structure of the 50 nm Pb$_{1-x}$Sn$_{x}$Se (x=0.21) can now be calculated using an envelope function method\cite{bastard1990wave}, previously used in Refs. \citen{krizman2018dirac} and \citen{wang2020weak}. Following the basis and model shown in Ref. \citen{wang2020weak}, the k.p matrix Hamiltonian of Pb$_{1-x}$Sn$_{x}$Se, is utilized for this operation, with $k_z=-i\partial_z$. The energy gap of the TCI well is fixed at $E_g=-30$ \si{meV} at low temperature. The energy gap in the barrier is fixed to 2 \si{eV}. Both the substrate and the capping layer have a wide energy gap exceeding 1 \si{eV}, however, we have checked that varying energy gap of the barriers in the model above 1 \si{eV} has a minimal impact on the band structure of the Dirac states of the material. The band offsets of the conduction and the valence bands are assumed equal and opposite in sign to account for the band inversion that takes place at the interface.
 
\begin{figure}
    \centering
    \includegraphics{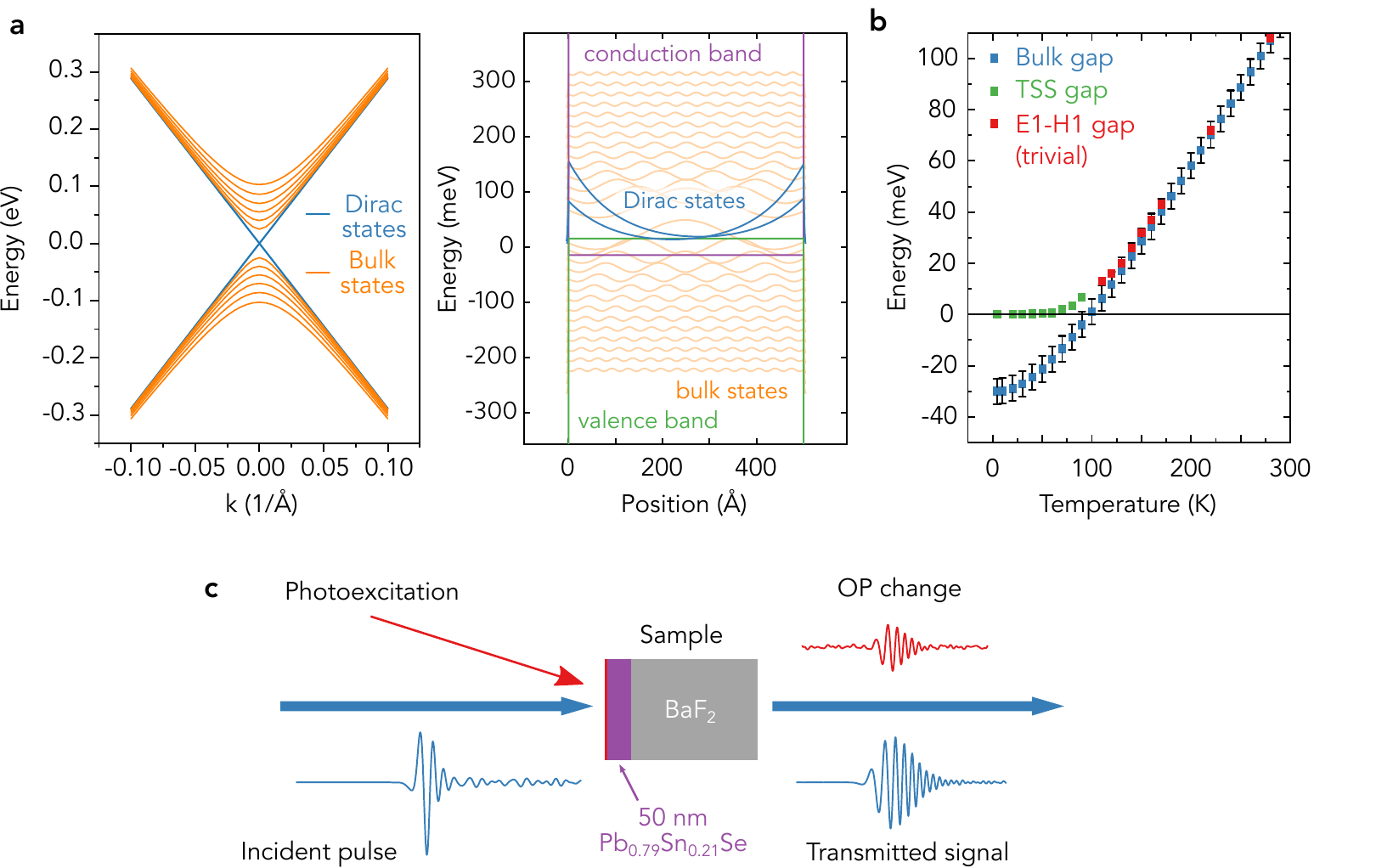} 
    \caption{\label{fig:sampleinfo}System and Pb$_{1-x}$Sn$_{x}$Se (x=0.21) sample information. a. Calculated band diagram of the Pb$_{1-x}$Sn$_{x}$Se (x=0.21) film at 4.2 K (left), and zone center ($k_t=0$) wavefunctions (right). b. Gap information of Pb$_{1-x}$Sn$_{x}$Se (x=0.21) versus temperature. The bulk bandgap is represented by the blue squares, the bandgaps of the topological surface states (TSSs) are represented by green squares, and the red squares represent the E1 gap (the gap of the lowest subband when trivial). c. Simplified version of the optical-pump terahertz-probe spectroscopy (OPTP) measurement.}
\end{figure}

 We fix the k.p matrix along the growth direction to $\hbar \nu_z=2.6$ \si{eV\angstrom} and we assume the bulk valleys to be isotropic so that $\nu_z=\nu_x$, as typically seen for Pb$_{1-x}$Sn$_{x}$Se with x>0.20.\cite{krizman2018dirac} The model yields the band dispersion shown in Fig. \ref{fig:sampleinfo}a, with gapless Dirac surface states dispersing the energy gap formed by the quantum well subbands. To model the temperature dependence of the band structure, we use the energy dependent gap empirically determined from  Equation \ref{e1}, a repeat of our calculations with $E_g (0.21,T)$ implemented in the model for various T. The model yields the energy gap of the topological surface states (TSS) in the topological regime as it evolves from 0 \si{meV} up to a few meV when $E_g (0.21,T)$ is small but negative. When $E_g (0.21,T)$ becomes positive, the model is equivalent to that of a conventional narrow gap semiconductor and yields the energy separation between the top most hole subband H1 and bottom most electron subband E1. The evolution of the TSS gap and that of E1-H1 gap is shown in Fig. \ref{fig:sampleinfo}b. 

As can be seen in Fig. \ref{fig:sampleinfo}b, a topological phase transition is seen to happen at 110$\pm$15 K (the uncertainty is due to the propagated uncertainty of $E_g$). When the temperature goes lower than this point, the topological surface states (TSS) with a gapless Dirac point appears, and the carrier in those states become so-called Dirac fermions. With our knowledge of the bulk gap, we can compute the band dispersion of the Dirac states and its gap acquisition through transition. The unique massless Dirac fermions in the surface states are the key to understand the dynamics present in this material. In order to study those dynamics, OPTP is used. The principle of this system can be seen in Fig. \ref{fig:sampleinfo}c: a high-power short-pulse laser is used to excite both a photoconductive antenna (PCA) and the sample, and the terahertz generated by the PCA is passed through the sample and modified by the photoexcited carriers. The terahertz signal is then detected using electro-optic sampling in a ZnSe crystal and balanced detection. For this measurement, the pulse used to excite the sample was at 800 nm (1.55 eV) with a duration of 86 fs. Because we were limited by the high repetition rate of our Ti:Sapphire laser, the optical pump beam’s fluence at the sample was limited to 1.6 \si{\mu J/cm^2} for all measurements. The bandwidth of our system is about 6 THz, and the SNR is about 60 dB. A dual lock-in amplifier scheme is used to remove long-term drift: the pump beam is chopped, and two lock-in amplifiers are used to measure the transmitted terahertz. One represents the difference in terahertz with the pump on versus the pump off, while the other represents the average. These two signals can be combined to produce the spectrum with the pump on versus the pump off.

\begin{figure}
    \centering
    \includegraphics{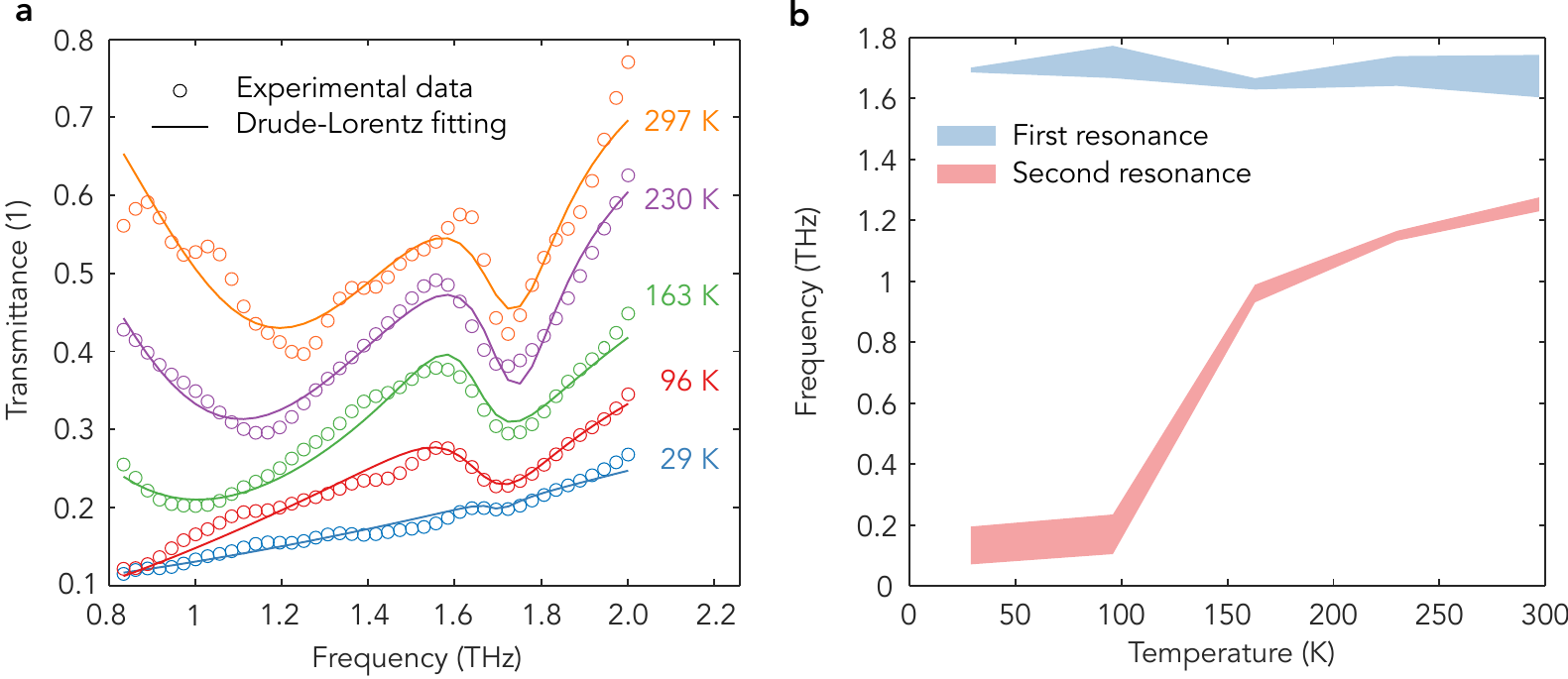}
    \caption{ \label{trans}a. Pb$_{1-x}$Sn$_{x}$Se (x=0.21) transmittance at different temperatures. Measurements are indicated by circles and the Drude-Lorentz fittings are indicated by lines. b. Resonance frequencies as a function of temperature. The widths of the curves represent one standard deviation.}
\end{figure}

Using the time-domain system, we first measured the transmittance of the Pb$_{1-x}$Sn$_{x}$Se (x=0.21) sample as the temperature is changed. As shown in Fig. \ref{trans}a, we focus on five temperature points, two of which (29 K and 96 K) are well below the temperature point at which there exist topological surface states, and three that are above (163 K, 230 K and 297 K). The transmittance is calculated by comparing the signal through a substrate without the TCI to one that is measured with the TCI. Due to the lower SNR at the two sides of the spectrum, we select only the highest SNR region from 0.7 THz to 2 THz for analysis---below the Reststrahlen band of the BaF$_2$ substrate---and fit the transmittance data using a Drude-Lorentz model with two resonant frequencies. Two pronounced resonances are observed: one at 1.7 THz (7 meV) that is relatively constant in temperature, and one lower-frequency branch that is strongly temperature-dependent, below 1.2 THz (5 meV). Neither can be attributed to direct electronic transitions, as the energies are too low to be explained by intersubband absorption ($\Delta E>14$ \si{meV} for all occupied subbands). We attribute the constant-temperature resonance as arising from weak coupling to bulk TO phonons, as this frequency occurs close to the value previously measured by FTIR spectroscopy (50 \si{cm^{-1}})\cite{novikova2018infrared}. The strongly temperature-dependent resonance can perhaps be best explained as coupling to shear vertical surface modes, which were previously found (in Pb$_{0.7}$Sn$_{0.3}$Se) to have energies that strongly depend on the emergence of Dirac fermions, shifting to low energies at temperatures below the transition \cite{kalish2019contrasting}. This effect arises as a result of the strong electron-phonon interaction but requires further study.

\begin{figure}
    \centering
    \includegraphics{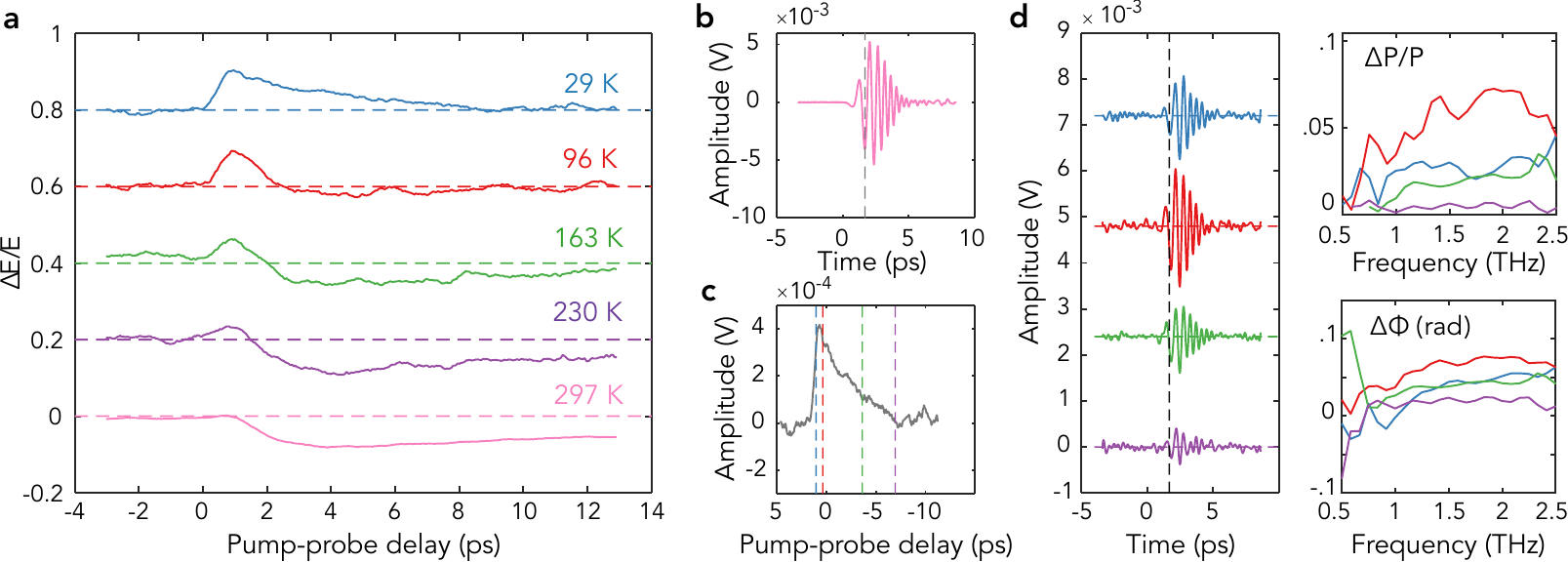}
    \caption{ \label{OP}a. Effect of an optical pump on the amplitude of the transmitted terahertz pulse at different temperatures, offset for clarity. (The dashed lines indicate the offset for each curve.) An increase in transmission is notable at low temperatures, while only a decrease is evident at high temperatures. b. Transmitted TDS pulse shape of Pb$_{1-x}$Sn$_{x}$Se at 29 K. c. OP pulse-induced difference of THz field through Pb$_{1-x}$Sn$_{x}$Se at 29 K as a function of pump-probe delay. d. Dual lock-in difference signal measured at different time points marked in c, in both time (left) and frequency (right) domain.}
\end{figure}

After the transmittance spectra of the Pb$_{1-x}$Sn$_{x}$Se sample were obtained using normal time-domain spectroscopy, OPTP measurements were performed. By locating the first stage of the TDS signal at the point marked by the black dashed line in Fig. \ref{OP}b and scanning the delay of the high-power infrared pump pulse, information about the time-resolved carrier dynamics could be measured, as is shown in Fig. \ref{OP}a. Figs. \ref{OP}b, \ref{OP}c, and \ref{OP}d shows the time scale of the detailed OP signal. In Fig. \ref{OP}c, one can see that the transmitted field reaches a maximum at a pump-probe delay of 1.6 ps (with a 1 ps rise time), and then decays over the next 6 ps. In order to investigate the frequency response, the change in the transmitted signal induced by the pump was plotted at several pump-probe delays. Each curve in Fig. \ref{OP}d represents the transmitted pulse at the different time points marked with the same color dash line in \ref{OP}c. These are also converted into the frequency domain. Note that the optical pump does not induce a strong frequency dependence in the transmission, as is expected of the optical properties of Dirac cones. 

Fig. \ref{OP}a illustrates the photoexcited carrier dynamics as a function of temperature. At room temperature (T=297 K), the transmittance of the sample decreases as the pump pulse arrives at the sample. This is a well-explained phenomenon that is justified by increase of the number of free carriers. At this temperature point, the sample has a bandgap around 115 meV, which makes its behavior similar to conventional semiconductors: only massive fermions are present, the gapless Dirac cone doesn't exist, and due to the increase of the free carrier concentration, the transmittance decreases after photoexcitation. The induced photoconductivity at room temperature decays at a rate of about 27 ps due to interband scattering. However, at low temperatures (T=29 K), an increase in transmittance after the sample is excited is observed. This is an anomalous phenomenon that has also been observed in some well-explored 2D materials, such as Bi$_2$Se$_3$ and graphene.\cite{shicontrolling2014,hafez2017effects, jnawaliobservation2013, lunegative2017, valdesaguilartime-resolved2015,tielrooij2013photoexcitation}. A decay time of about 6 ps is observed. A combination of positive and negative photoconductivity is observed at different pump delay positions at moderate temperatures (T=96, 163, 230 K) which can be explained by the coexistence of the Dirac fermions from the gapless surface states and the massive fermions from the bulk states. We also find that the strength of the pump-induced signal is also changing with temperature. As the temperature is lowered, the effect of the massive electrons is reduced, as the low fluence ensures that the incident light is screened by the Dirac fermions.

To further explore the anomalous transmittance increase at low temperatures, we calculate the photoconductivity of the sample at 29 K using\cite{wu2013sudden}
\begin{equation} \label{e3}
   T(\omega) =\frac{1+n}{(1+n+Z_{0}G(\omega)) e^{i\phi}},
\end{equation}
where $T(\omega)$ is the complex transmittance of the sample, n is the refractive index of the substrate, $Z_{0}$ is the impedance of free space, $G\left(\omega\right)$ is the conductivity, and $\phi$ is the phase difference introduced by the thickness difference of the substrate and the Pb$_{1-x}$Sn$_{x}$Se sample.

\begin{figure}
    \centering
    \includegraphics{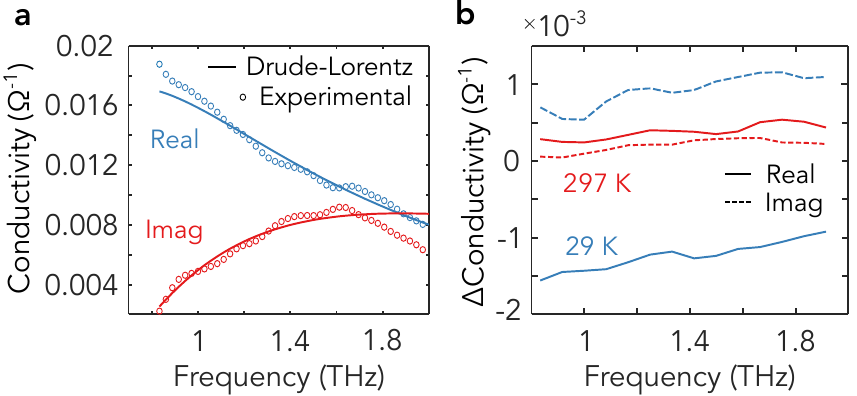}
    \caption{ \label{fit} a. Real and imaginary part of the conductivity at 29 K with no optical pump. b. Real and imaginary part of the photoinduced conductivity at 29 K and 297 K.}
\end{figure}

Fig. \ref{fit}a shows the calculated real and imaginary part of the conductivity as a function of frequency for the unpumped sample. The pumped sample is very similar, so in Fig. \ref{fit}b we plot the differential conductivity. To better understand the physical origin of the photoinduced conductivity here, the conductivity is fitted with a two-term Drude-Lorentz model, including a Drude term, a Drude-Lorentz term (which modelling the TO phonon resonance), and a frequency-independent contribution to the dielectric constant (which accounts for the effect of higher-energy excitations):\cite{valdes2013aging,sim2014ultrafast}
\begin{equation}\label{e4}
   G_j(\omega) =\left(-\frac{\omega_{pD}^2}{i\omega-\Gamma_{D}}-\frac{i\omega\omega_{pDL}^2}{\omega_{DL}^2-\omega^2-i\omega\Gamma_{DL}}-i\left(\epsilon_{\infty}-1\right)\omega\right)\epsilon_{0}d,
\end{equation}
where $\omega_{pD}$ is the plasma frequency of the Drude oscillator, $\Gamma_{D}$ is the scattering rate of the Drude oscillator, $\epsilon_{\infty}$ is the dielectric constant at high frequency, d is the sample thickness, $\omega_{pDL}$ and $\Gamma_{D}$ are the parameters for the Lorentz oscillator. 

The origin of the negative photoinduced conductivity can be understood by considering the Drude conductivity in the low-frequency limit. The low-frequency conductivity dynamics are governed by the Drude weight $D_{Drude}=\frac{d\omega_{pD} ^2}{(2\pi)^2}$ and the scattering rate $\Gamma_{Drude}$, which is related to the carrier concentration and the mobility respectively. The real part of the conductivity is determined by the ratio of the Drude weight and the scattering rate. For conventional semiconductors with bandgaps, the real part of photoinduced conductivity typically increases since the photoinduced carriers increase the Drude weight. For Dirac cones, this effect is typically offset by an increase in scattering rate, as the photoexcited electrons become hot \cite{sim2014ultrafast}.
At room temperature, the Pb$_{1-x}$Sn$_{x}$Se sample possesses only massive states, which makes it behave like conventional semiconductors. However, as lower temperature, the surface states become gapless and the increase of scattering rate becomes dominant, leading to a negative photoinduced conductivity.

To confirm that the scattering rate increases, we fit the complex conductance spectra to the same Drude-Lorentz model with and without the pump. While the Drude weight $D_{Drude}$ of the pumped Pb$_{1-x}$Sn$_{x}$Se (x=0.21) sample increased by 2.1\%, consistent with a small increase in the density of free carriers, a much larger increase of the scattering rate, $\Delta\Gamma_{D}$=47.8\%, is observed. As a result, the ratio $\frac{D_{Drude}}{\Gamma_{Drude}}$ acutally decreases. This causes the real part of the photoinduced conductivity, $\delta G\equiv{G_{pumped}-G_{unpumped}}$, to be negative as is shown in Fig. \ref{fit}b. This explains the anomalous increase of the signal after the photoexcitation. Therefore, the 6 ps decay rate of the negative photoconductivity should be interpreted as an intrasubband (i.e., intra-Dirac cone) relaxation rate, not as a lifetime of the Dirac fermions.

\section{Conclusion}
In conclusion, we have used optical-pump terahertz probe spectroscopy to characterize TCI Pb$_{1-x}$Sn$_{x}$Se heterostructures. At low temperatures where the topological transition happens, one observes a negative photoinduced conductivity immediately after the photoexcitation, a result of the generation of hot surface state Dirac fermions. The relaxation of these Dirac fermions is relatively fast, occurring in about 6 ps. At room temperature, only normal massive carriers are observed, and the transmittance behaves like a normal photo-excited semiconductor, with a carrier lifetime of about 27 ps. At moderate temperatures, both carriers are observed. Our observations provide critical details for potential applications of Pb$_{1-x}$Sn$_{x}$Se and provide a direct measurement of the topological character of Pb$_{1-x}$Sn$_{x}$Se thin films, which will be of use for the characterization of more sophisticated bandstructure-engineered heterostructures.

\begin{acknowledgement}


All authors acknowledge support from the NDnano seed grant program. D.B. and Z.X acknowledge support from the AFOSR under grant no. FA9550-20-1-0192, from the NSF under ECCS-2046772, and from ONR under N00014-21-1-2735. BA.A, J.W and X.L acknowledge support from NSF1905277 and DMR-1905277.

\end{acknowledgement}


%

%


\bibliography{ref}

\end{document}